\documentclass[aps,,12pt]{revtex4}
\def\lsim{\raise0.3ex\hbox{$\;<$\kern-0.75em\raise-1.1ex\hbox{$\sim\;$}}}
\def\gsim{\raise0.3ex\hbox{$\;>$\kern-0.75em\raise-1.1ex\hbox{$\sim\;$}}}

 \normalsize
\usepackage{graphicx,graphics}
\begin{document}

\title{Can we have another light ($\!{\bf\sim 145}$ GeV) Higgs boson?}
\author{S. Khalil$^{1,2}$  and   S. Moretti$^{3,4}$}
\affiliation{$^{1}$ Center for Fundamental Physics, Zewail City {of} Science and Technology,
6 October  City, Giza, Egypt.\\
$^{2}$ Department of Mathematics, Faculty of Science,  Ain Shams University, Cairo, Egypt.\\
$^{3}$ School of Physics \& Astronomy, University of Southampton, Highfield, Southampton, UK.\\
$^{4}$ Particle Physics Department, Rutherford Appleton Laboratory, Oxon OX11 0QX, UK.}
\date{\today}

\begin{abstract}
A second light Higgs boson, with mass of approximately 145 GeV,  is predicted by non-minimal Supersymmetric models. 
This new particle can account for an apparent $\sim 3 \sigma$ excess recorded by the CMS experiment at the Large Hadron Collider (LHC) during Run 1.
We show how this can be explained in a particular realisation of these scenarios, the $(B-L)$ Supersymmetric Model (BLSSM), which also has other captivating features, like offering an explanation for neutrino masses and relieving the small hierarchy problem of the Minimal Supersymmetric Standard Model (MSSM).
\end{abstract}

\maketitle

The paradigm that particle physics is minimalist in Nature, at least as far as Electro-Weak Symmetry Breaking (EWSB) goes \cite{Wilczek}, may simply  be the result of early appearance if one realises that the Standard Model (as we know it) appeared to be so initially, but then it revealed itself rather more articulated than we  thought or hoped. As for its interactions, there was first the photon, uncharged and massless. We later discovered its massive weak companions, $Z$ and $W^\pm$, the latter being charged too. Even the gluon is one {of} eight, actually.  Concerning matter, {the story started with} one generation of quarks and leptons/neutrinos, that was sufficient to keep our world stable. Then somebody ordered the muon \cite{Rabi} and all apparently fell apart. With it also came its neutrino (not that we saw it at the time or even now). Strangely yet charmingly, the quarks were no less zealous, producing their own second generation of two offsprings within a few decades.   
Discovering the third generations of fermions was even more upsetting, so bulky in comparison to the preceding two.

We may then have to dismiss a minimalist attitude also for the Higgs sector, eventually. 
Intriguingly, in fact, in the search for Higgs bosons during Run 1, the CMS collaboration also found potential signals for another Higgs boson, $h'$,  with mass $m_{h'}\ge 136.5$ GeV in the ${h'}\to ZZ \to 4l$ decay mode \cite{Chatrchyan:2013mxa}, wherein a $\sim2\sigma$ excess is appreciable in the vicinities of 145 GeV, 
and in the ${h'} \to \gamma \gamma$ decay channel, wherein the local $p$-value indicates possibly significant excesses very near both
137 and 145 GeV at the  $\sim 2.9$ and $\sim2 \sigma$ level, respectively \cite{CMS:2013wda}. Also, various anomalies for a mass $\gsim137$ GeV or above have emerged in several other channels, from both ATLAS and CMS at the LHC as well as both CDF and D0 at the Tevatron \cite{anomalies}.

We ought to be prepared this time, then, for a non-minimal Higgs sector. Supersymmetry (SUSY), for example, calls for it. We are rather fond of this ultimate (potential) symmetry of Nature in fact, as for the first time a theory would probably solve more problems that it could create. In particular,
SUSY, whichever shape or form of it is actually realised in Nature, wants a light Higgs boson, with mass similar to that of the weak gauge bosons. (No such a claim can be made by the SM instead.)  Alas, SUSY has not been seen, yet. While disturbing per se, this fact may actually be a consequence of (yet again) a flawed approach, that assumes that SUSY is also minimal. Just like the SM actually is not for most of its parts (and we claim it to be for none), SUSY needs not be so either. Unsurprisingly, if one dismisses minimalism in SUSY, one may find at the same time an explanation for the absence of its manifestations at present \footnote{Typical sparticle signals may undergo longer cascades in a non-minimal SUSY
model, thereby escaping traditional SUSY searches.} as well as a hint of where Higgs companions might be.   

It is remarkable though that an explanation for a second light Higgs particle cannot be found inits minimal version,
 the MSSM.  Although it has 
an additional neutral and CP-even Higgs boson, it cannot account for the possibility of the aforementioned possible double Higgs peak, with masses at $\approx125$ and $\approx145$ GeV. Furthermore, the solid experimental evidence for neutrino oscillations, pointing towards non-vanishing neutrino masses, hints at favouring other SUSY realisations than the MSSM.  
For example. a minimal extension of the latter is based on the gauge group $SU(3)_C \times SU(2)_L \times U(1)_Y \times U(1)_{B-L}$. In the ensuing model, the aforementioned BLSSM, the $(B-L)$ symmetry breaking is related to the soft SUSY breaking scale, {i.e.}, ${\cal O}(1)$ TeV.  In this class of models, with TeV scale right-handed neutrinos, light neutrino masses can be naturally implemented through the inverse see-saw mechanism \cite{Khalil:2015naa}. The particle content of this model includes the following Superfields in addition to those in the MSSM: two SM singlet chiral Higgs superfields $\chi_{1,2}$, with the Vacuum Expectation Value (VEV) of their scalar components spontaneously breaking  $U(1)_{B-L}$  and with $\chi_2$ necessarily required to cancel the $U(1)_{B-L}$ anomaly; three sets of SM singlet chiral superfields $N_i, S_{1_i}, S_{2_i} (i =1,2,3)$, to implement the inverse see-saw mechanism (also 
needed to cancel the $(B-L)$ anomaly). In the BLSSM with inverse see-saw, the so-called small hierarchy problem of the MSSM, wherein the discovered Higgs boson mass of 125 GeV is dangerously close to its predicted absolute upper limit (130 GeV or so), is  relieved by providing (s)neutrino mass corrections which can up-lift this value to 170 GeV or so \cite{Elsayed:2011de}. 

The gauge kinetic term induces mixing in the squared-mass matrix of the BLSSM CP-even neutral Higgs fields. In this regard, we can obtain a SM-like Higgs boson, $h$, with mass of order $125$ GeV, and the lightest $(B-L)$ neutral Higgs state, $h'$, with mass $m_{h'} \simeq {\cal O}(100$ GeV).  As shown in Ref. \cite{Abdallah:2014fra}, the lightest BLSSM-specific Higgs boson, $h'$,  can be the second lightest Higgs boson (with a mass $\sim 145$ GeV). (In addition, there are two heavy Higgs bosons, $H$ and $H'$, with masses of order TeV.)

\begin{figure}[t]
\centering
\includegraphics[scale=0.75,angle=0]{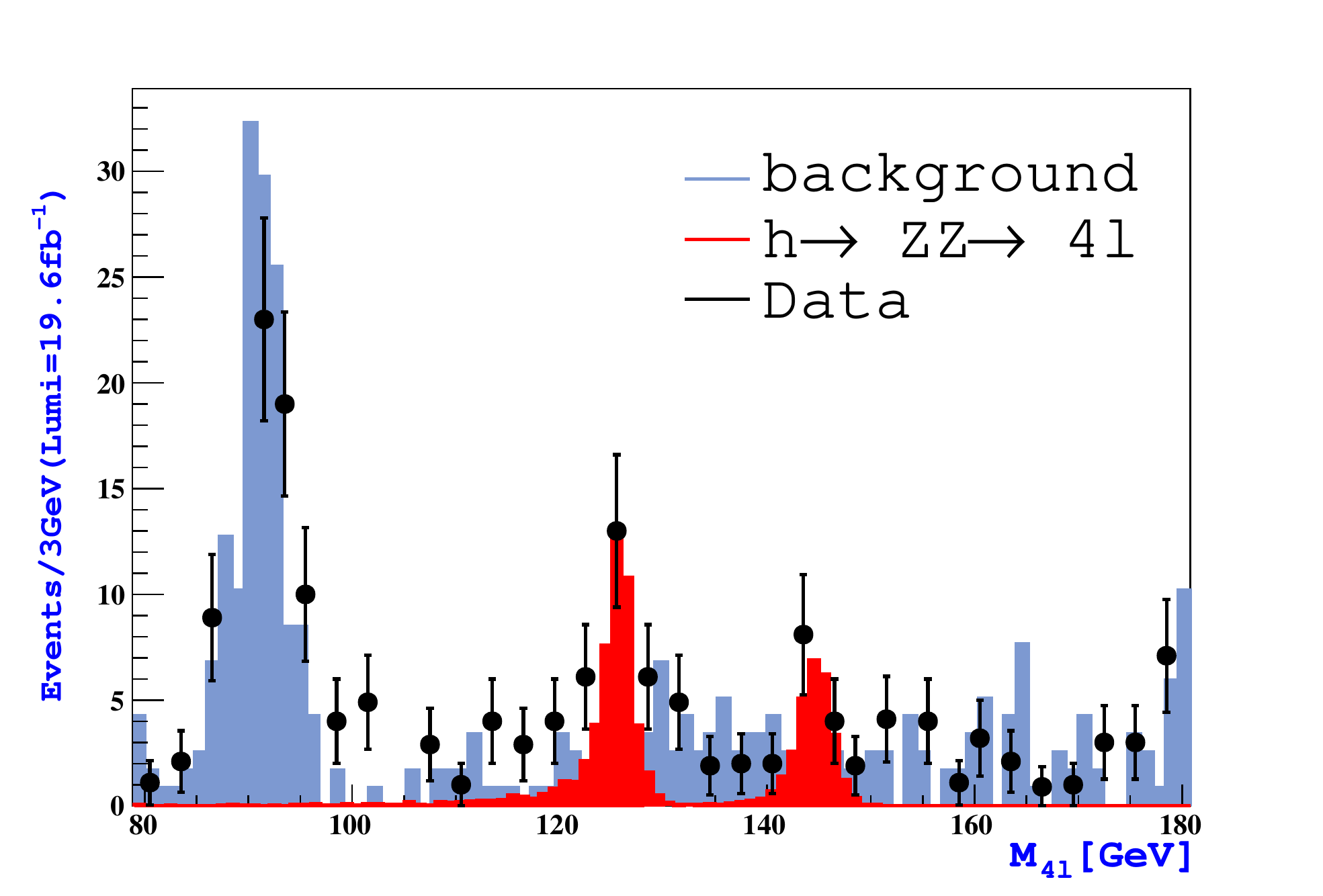}
\caption{The number of events from the signal 
$pp  \to  h,h' \to  ZZ \to  4l$ (red filled histogram) and
from the background $pp \to Z \to 2l\gamma^*\to 4 l$ (blue filled histogram)
 versus the invariant mass of the outgoing particles (4-leptons) against data taken from \cite{Chatrchyan:2013mxa}.}
\end{figure}

In Fig.~1 we show the invariant mass of the 4-lepton final state from $ pp \to h' \to ZZ \to 4l$ 
as obtained at $\sqrt{s}={8}$ TeV after 19.6 fb$^{-1}$ of luminosity, after applying a $p_T$ cut of 5 GeV on the four leptons. The SM model backgrounds from the $Z$ and {125} GeV Higgs boson decays,  $pp \to Z\to 2l \gamma^* \to 4 l$ and $pp \to h \to ZZ \to 4l$, respectively, are taken into account, as
demonstrated by the first two peaks in the plot (with the same $p_T$ requirement). It is clear that the third peak at $m_{4l} \sim 145$ GeV, produced by the decay of the BLSSM Higgs boson $h'$ into $ZZ \to 4 l$, can reasonably well account for the events observed by CMS \cite{Chatrchyan:2013mxa} with the 8 TeV data.

\begin{figure}
\centering
 \includegraphics[scale=0.85,angle=0]{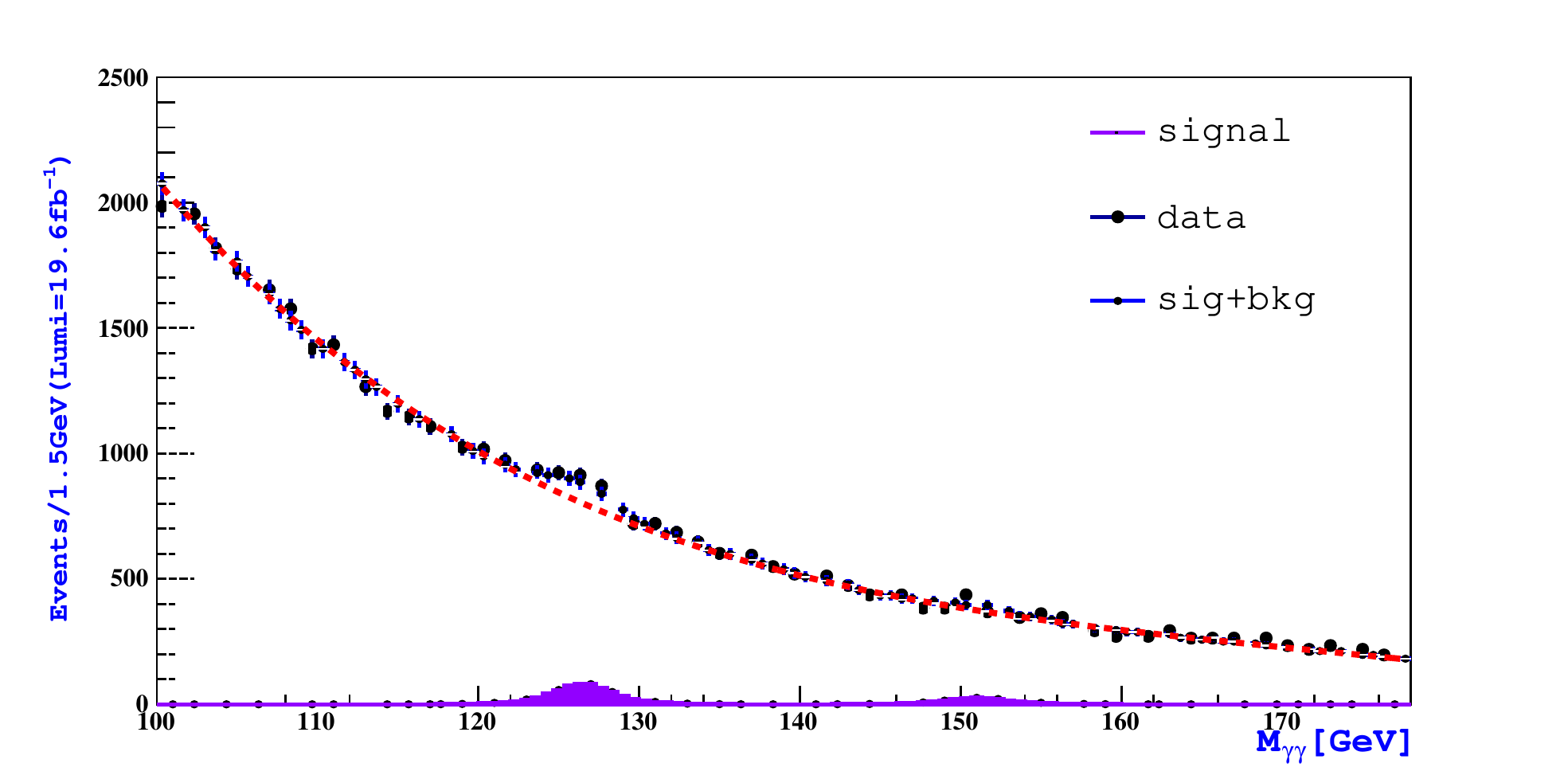}
 \caption{The number of events from the signal  $pp  \to  h,h' \to  \gamma \gamma $ (purple filled histogram), from the background $ pp \to \gamma \gamma$ (red dotted histogram) and from
the sum of these two
 (blue dotted histogram) versus the invariant mass of the outgoing particles (di-photons) against data taken from \cite{CMS:2013wda}.}
\end{figure}

The distribution of the di-photon invariant mass is presented in Fig.~2, again for a centre-of-mass energy of $\sqrt{s}={8}$ TeV and luminosity of 19.6 fb$^{-1}$. As previously, here too, the observed $h\to \gamma\gamma$ SM-like signal around 125 GeV is taken as background (alongside the continuum) while the (rather subtle) $Z\to \gamma\gamma$ background can now be ignored \cite{Moretti:2014rka}. As expected, the sensitivity to the $h'$ Higgs boson is severely reduced with respect to the presence of the already observed Higgs boson, yet a peak is clearly seen at 145 GeV or so and is very compatible with the excess seen by CMS \cite{CMS:2013wda}.

Before closing, we should also mention that the $h'\to\gamma\gamma$ enhancement found in the BLSSM may be mirrored in the $\gamma Z$ decay channel \cite{Hammad:2015eca} for which, at present, there exists some constraints, albeit not as severe as
in the $\gamma\gamma$ case. 

Doubtless, if the two peaks are real, an explanation for these can be found in other SUSY models than just the BLSSM \footnote{The details of the BLSSM parameter point used to fit the CMS data are available upon request.}. Notable is the one given
by the Next-to-MSSM (NMSSM) \cite{2higgses}, wherein the non-minimality is achieved by enlarging further the MSSM Higgs spectrum iself, by an additional Higgs singlet state, to elegantly obviate the so-called `$\mu$-problem of the MSSM \cite{Ellwanger:2009dp}, another of its drawbacks. Other explanations can well be found within SUSY for the possible existence of a second peak. Far for judging the relative merits
of each of these (and possibly other) SUSY scenarios, our intention was merely to alert the community that the hunt for Higgs bosons if far from over and that there
may already be hints from data on where to find the next one. 

It is now wait and see for a few more months. The reward could be tantalising: not only the confirmation of a second scalar peak (i.e., of a non-minimal Higgs sector) but also in a mass region that would provide evidence (albeit circumstantial) of SUSY in a non-minimal form. As far as we are concerned, in fact, despite the potential benefits of a widespread Occam's razor attitude,  
we cannot disagree with the fact that ``everything should be made as simple as possible, but not simpler'', our favoured paraphrase of the Einstein's razor instead \cite{Einstein}.

\section*{Acknowledgements}
{We would like to thank A. Hammad for his valuable help in producing the plots.} 
The work of SK is partially supported by the ICTP grant AC-80 while SM's is through the NExT Institute and the
STFC Consolidated Grant ST/J000396/1. 
 The work of both SK and SM is also funded through the grant H2020-MSCA-RISE-2014 no. 645722 (NonMinimalHiggs).



\begin{thebibliography}{1}

\bibitem{Wilczek} F. Wilczek, 
Nature {\bf 496}, 439 (2013).

\bibitem{Rabi} I.I. Rabi, ``Who ordered that?'' (a quip in 1957, verbal).

\bibitem{Chatrchyan:2013mxa} 
CMS Collaboration,
  Phys.\ Rev.\ D {\bf 89}, 092007 (2014).

\bibitem{CMS:2013wda} 
  CMS Collaboration,
CMS-PAS-HIG-13-001 (2013) and  
CMS-PAS-HIG-13-016 (2013).
%


\bibitem{anomalies} CMS Collaboration, CMS-PAS-HIG-13-044 (2013);
ATLAS Collaboration, ATLAS-CONF-2013-012 (2013) and
Phys. Lett. B {\bf 726}, 88 (2013)
[Erratum, ibidem {\bf 734}, 406 (2014)];
CDF \& D0 Collaborations, Phys. Rev. D {\bf 88}, 052014 (2013).


\bibitem{Khalil:2015naa}
  S.~Khalil and S.~Moretti,
  arXiv:1503.08162 [hep-ph].

\bibitem{Elsayed:2011de}
  A.~Elsayed, S.~Khalil and S.~Moretti,
  Phys.\ Lett.\ B {\bf 715}, 208 (2012).

\bibitem{Abdallah:2014fra} 
  W.~Abdallah, S.~Khalil and S.~Moretti,
  Phys.\ Rev.\ D {\bf 91}, 014001 (2015).

\bibitem{Moretti:2014rka} 
  S.~Moretti,
  Phys.\ Rev.\ D {\bf 91}, 014012 (2015).

\bibitem{Hammad:2015eca} 
  A.~Hammad, S.~Khalil and S.~Moretti,
  arXiv:1503.05408 [hep-ph].

\bibitem{2higgses}
G. Belanger, U. Ellwanger, J. F. Gunion, Y. Jiang, S. Kraml,
arXiv:1208.4952 [hep-ph].
%

\bibitem{Ellwanger:2009dp} 
  U.~Ellwanger, C.~Hugonie and A.~M.~Teixeira,
  Phys.\ Rept.\  {\bf 496}, 1 (2010).

\bibitem{Einstein}
A. Einstein, ``On the Method of Theoretical Physics'', The Herbert Spencer Lecture, delivered at Oxford (10 June 1933),  published in Philosophy of Science, Vol. {\bf 1}, No. 2 (April 1934), pp. 163-169, see p. 165.

\end{thebibliography}
\end{document}